# Signatures of Anderson localization and delocalized random quantum states

Giorgio J. Moro, Giulia Dall'Osto, Barbara Fresch

Dipartimento di Scienze Chimiche, Università degli Studi di Padova

Via Marzolo 1, 35131, Padova, Italy

*Abstract*

We consider the notion of equilibration for an isolated quantum system exhibiting Anderson localization. The system is assumed to be in a pure state, i.e., described by a wave-function undergoing unitary dynamics. We focus on the simplest model of a 1D disordered chain and we analyse both the dynamics of an initially localized state and the dynamics of quantum states drawn at random from the ensemble corresponding to the minimum knowledge about the initial state. While in the former case the site distribution remains confined in a limited portion of the chain, the site distribution of random pure state fluctuates around an equilibrium average that is delocalized over the entire chain. A clear connection between the equilibration observed when the system is initialized in a fully localized state and the amplitude of dynamical fluctuations of a typical random pure state is established.

## 1. Introduction

Quantum localization has been drawing intense interest both because it represents a challenge to the principles of quantum statistical thermodynamics, and because of its possible implications for the emergence of novel technologies and quantum devices, that are protected against decoherence and can operate even at finite temperature. The quantum concept of localization was first introduced by Anderson[1] and in its original formulation refers to noninteracting particle in a random potential. The aspects of Anderson localization that are of interest to us and that will be investigated in the following, are related to the issue of equilibration and thermalization in isolated quantum systems. Recent experiments allowed the observation of the dynamics of mesoscopic quantum systems virtually isolated from the environment.[2-6] Because of the finiteness of the system and possible localization phenomena, the standard setting of quantum statistical mechanics based on the canonical density matrix and master equations is not suitable to describe the system dynamics. The entire system is in a quantum mechanical pure state and therefore the same notions of *equilibration* and *thermalization* needs to be revised.[7-10] The key point is that equilibration and thermalization are not necessarily equivalent. The dynamics is time-reversal invariant and for finite systems recurrent so, a priori, it seems far from clear in what sense equilibrium can be reached dynamically. However, one can always define an equilibrium average of any observable as its asymptotic time average. Much work has been done to show that under physically reasonable conditions, local observables that are initialized to a different value, evolves toward their equilibrium average and remain close to it for most of the time.[7-9, 11] A more complete description is obtained through the analysis of the entire distribution of the observables along an infinite time trajectory.[12-14]

A different and separate issue is to define whether and under which conditions the equilibrium average can be identified with a thermal average, where the thermal average is commonly referred to the traditional canonical or microcanonical density matrix of standard statistical quantum mechanics. Also for this second problem, different theoretical approaches have been developed as the *eigenstate thermalization hypothesis*[15-17] and the analysis of *typicality* in suitable ensemble of quantum pure states.[18-20] These works pointed out that relaxation of local observables toward a thermal value is a rather general process and few recent experiments nicely confirm this expectation.[4, 6] However, Anderson localized systems are evident exceptions to this scenario.

In this paper, we present a statistical analysis of the simplest model of a finite system exhibiting Anderson localization, that is a single particle in a linear chain whose sites are characterized by disordered potential. If the system is initialized in a localized site, the initial wave-packet evolves under the effect of the disordered Hamiltonian and so does the corresponding site occupation probability density. After the initial transient, the site population distribution attains its long time asymptotic profile, up to dynamical fluctuations. It is this process that we call equilibration dynamics and we emphasize that the equilibrium regime that is attained is dynamical because of fluctuations. The localization can be then identified according to the confinement of the equilibrium average occupation probability within a limited portion of the chain. Clearly, such an equilibrium state does not conform to thermodynamic expectation that would predict a nearly uniform distribution of site populations.

In section 2 and 3 we investigate this process of equilibration from an initially localized state both analytically and numerically and we establish the relation between the localization length defined by the equilibrium site distribution and the disorder strength of the chain. We mention here, and we will further discuss in section 3, that our analysis is based on the dynamics of the site population density, and this is determined by many energy eigenstates. From this point of view, the problem of localization shows many analogies with the problem of quantifying the extent of phase-space explored during intramolecular quantum dynamics.[21-24] This is different from the more traditional point of view on Anderson localization that is based on the localization properties of eigenstates.[25]

In section 4 we take a different approach and we investigate the statistical properties of random pure states for the same Anderson chain. An ensemble of random pure state is defined as a uniform distribution on the possible wave-functions in the Hilbert space with the only constraint of normalization.[18] In sharp contrast to the dynamics of a localized initial state, site populations of typical random pure states undergo fluctuating dynamics around a nearly uniform distribution along the chain, i.e. they are delocalized. The main result of this work is to establish a relation between the average localization length of the site populations realized when the system is initially prepared in a specific site and the fluctuations amplitude of the site population in a general quantum state drawn at random from the ensemble corresponding to the minimum knowledge about the initial state.[26,27] We argue that in general, the variability of the expectation value of any observable in a finite isolated quantum system has two sources of different nature: the fluctuations stemming from the time evolution of the wave-function[28] and a statistical variance due to the uncertainty in preparing the initial state. We show that these two variances are not independent[29] and we discuss the connection between their relative weight and the phenomenon of localization.

## 2. Equilibration dynamics and localization

We shall study the Anderson model[1] in its simplest formulation of a tight-binding Hamiltonian for the one-dimensional chain of length $N$ with periodic boundary conditions.[30] By denoting with $|j\rangle$ the quantum state localized at the $j$-th site, the matrix elements of the Hamiltonian read

$$H_{j,j'} := \langle j|\hat{H}|j'\rangle = \varepsilon_j \delta_{j,j'} + V\delta_{j,j'+1} + V\delta_{j,j'-1} \tag{1}$$

for $j, j' = 1, 2, \cdots, N$. Because of periodic boundary conditions, the site index has to be considered as a cyclic variable, so that $j$ is equivalent to $j \pm N$. The parameter $V$ is the unique hopping coefficient of the model, while the site energies $\varepsilon_j$ are independently and randomly chosen according to a given normalized distribution $p(\varepsilon)$. The degree of confinement attained with the time is controlled by the disorder of site energies which can be quantified by their variance,

$$\sigma_\varepsilon^2 := \int d\varepsilon\, p(\varepsilon)\varepsilon^2 \tag{2}$$

having chosen the average as the origin of the energy scale, i.e., $\int d\varepsilon\, p(\varepsilon)\varepsilon = 0$. The results discussed in the following are obtained by using a Gaussian distribution parameterized by the standard deviation $\sigma_\varepsilon$. Thus, the controlling parameter is the ratio $\sigma_\varepsilon/V$ which corresponds to use the hopping coefficient $V$ as the energy unit in specifying the chain disorder. Then the limits $\sigma_\varepsilon/V \to 0$ and $\sigma_\varepsilon/V \to \infty$ define the conditions of minimum and maximum disorder of the chain, respectively.

From the numerical diagonalization of the matrix Eq. (1), one obtains the Hamiltonian eigenstates $|E_k\rangle$ and eigenvalues $E_k$

$$\hat{H}|E_k\rangle = E_k|E_k\rangle \qquad (3)$$

for $k=1,...,N$. In the limit $\sigma_\varepsilon/V \to 0$, when the coupling energy $V$ between neighbors dominates, a perfect lattice is recovered, and the eigenstates are extended over all the chain. This is valid for finite system size only, because localization theory asserts that in the limit of large systems all electron states are localized in the 1D case,[25] provided that $\sigma_\varepsilon/V \neq 0$. For sufficiently large $\sigma_\varepsilon/V$, when the chain disorder dominates, the eigenstates are strongly localized. The time evolution of the system wave-function $|\psi(t)\rangle$ is explicitly evaluated on the basis of the solutions of Eq. (3), once the initial condition $|\psi(0)\rangle$ is specified, namely

$$|\psi(t)\rangle = e^{-i\hat{H}t/\hbar}|\psi(0)\rangle = \sum_{k=1}^{N}|E_k\rangle e^{-iE_kt/\hbar}\langle E_k|\psi(0)\rangle \qquad (4)$$

The direct diagonalization of the Hamiltonian matrix Eq. (1) allows fast calculations of the quantum dynamics of chains with thousands of sites in a standard PC.

In this section, we shall analyze the delocalization dynamics arising from an initial fully localized state $|\psi(0)\rangle = |j_0\rangle$ at a given site $j_0$. The process of delocalization can be followed by monitoring the time evolution of the site populations

$$|\langle j|\psi(t)\rangle|^2 = \left|\sum_{k=1}^{N}\langle j|E_k\rangle e^{-iE_kt/\hbar}\langle E_k|j_0\rangle\right|^2 = \rho(j_0|j,t) \qquad (5)$$

which can be interpreted as the conditional probability $\rho(j_0|j,t)$ of finding the excitation at the $j$-th site if the initial location was $j_0$, with normalization $\sum_{j=1}^{N}\rho(j_0|j,t)=1$. Figure 1 shows the site distribution $\rho(j_0|j,t)$ as a function of the site displacement $\Delta j = j - j_0$ at selected times (the parameters of the model are specified in the figure caption). One observes an initial regime when the excitation initially localized at $j_0$ spreads over the neighboring sites (first and second panels in Fig. 1). After this initial transient evolution, the site populations still fluctuate, but the excitation shows a stable localization within a limited portion of the chain, the other sites having a negligible probability. Roughly speaking, one can recognize a localization length according to the portion of the chain having significantly populated sites after the transient evolution (third and fourth panels in Fig. 1). As we will discuss further below, the details of the time evolution of the site distribution also depend on the realization of the set of random site energies $(\varepsilon_1, \varepsilon_2, \cdots, \varepsilon_N)$ for the given energy distribution $p(\varepsilon)$. Other realizations of site energies would produce different site probabilities $\rho(j_0|j,t)$ but the general features of the dynamics described above are preserved in different realizations of the site energies for a fixed disorder strength $\sigma_\varepsilon/V$.

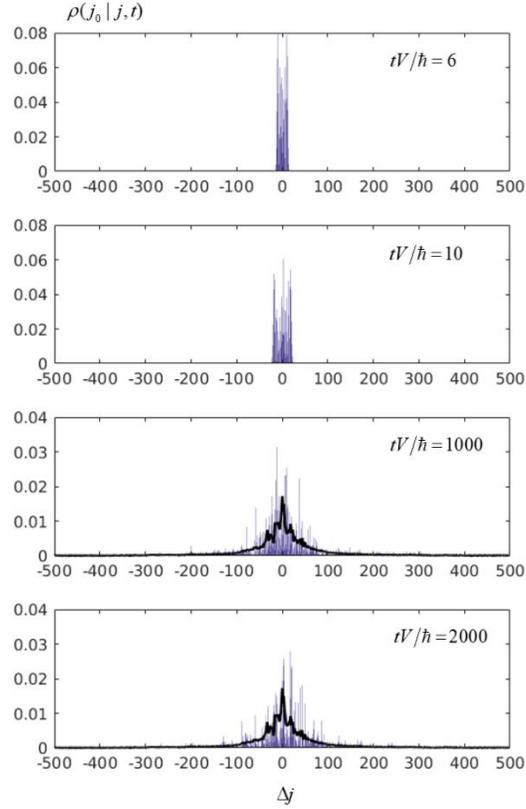

**Figure 1**. Time evolution of the site probability distribution $\rho(j_0 | j,t)$ as a function of the displacement $\Delta j = j - j_0$ from the initial site $j_0 = 1$. The panels show the site populations at increasing scaled times (from top to bottom) for a realization of the Anderson chain of length $N = 10^3$ with disorder strength $\sigma_\varepsilon / V = 0.24$. The equilibrium average distribution is also shown as a black trace in the bottom panels.

Let us analyze in more detail the time dependence of the probability distribution $\rho(j_0 | j,t)$ by focusing on the population of a specific site $j$, leaving aside for the moment the distribution within the chain. In Fig. 2 we have reported the time evolution for three sites within the localization length. One can recognize an initial regime (inset of Fig. 2) when the population of the initial state $j_0$ (blue line) decreases being transferred to neighboring sites while the population of other sites grow from the initial vanishing value. Afterwards a kind of stationary regime is attained with random fluctuations about a well-defined mean value. Because of the decomposition Eq. (4) for the wave function, the time dependence of a site probability $\rho(j_0 | j,t)$ is determined by the superposition of oscillatory contributions. However, given the considerable number of contributions with nearly random frequencies (differences between $E_k / \hbar$ pairs), such a superposition leads to a time profile resembling that of stationary fluctuations. The same behavior has been found in other finite dimensional quantum models[31, 32] when the Hamiltonian has, at least partially, a random character.

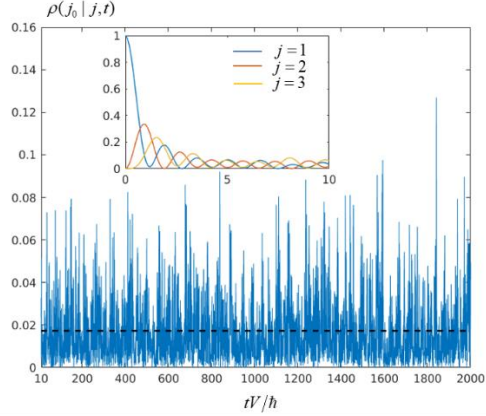

*Figure 2. Time evolution of selected site populations. In the inset, the short-time evolution of the probability $\rho(j_0 | j, t)$ of sites $j = 1, 2, 3$ for $j_0 = 1$ is shown. The model parameters are the same as in Figure 1. The main panel shows the long-time dynamics, after the initial transient, of the population of the first site $j = 1$, while the dashed line denotes the corresponding time average.*

The evolution of site probabilities, as displayed in Figs. 1 and 2, can be interpreted as an equilibration process[7, 8] due to the dynamics of the quantum pure state. By starting from an initial state prepared as the fully localized wave-function at a given site $j_0$, the system undergoes a delocalization process controlled, according to Eq. (5), by the evolution in the manifold of the expansion coefficients of the accessible Hamiltonian eigenstates $|E_k\rangle$ having a significant weight $|\langle E_k | j_0 \rangle|$. The final condition within such a manifold is that of an equilibrium having a dynamical character because of the coherent (oscillating) contribution of each eigenstate in Eq.(5). It should be stressed that such an equilibrium does not correspond to the thermodynamic equilibrium condition. Indeed, in correspondence of a thermalized state[33] one should expect an homogeneous distribution of the excitation along the chain. On the contrary, in the case of Fig. 1 equilibration leads to a site distribution that remains confined within a limited portion of the chain. In this sense, the Anderson's tight-binding Hamiltonian represents the simplest model for studying quantum equilibration without thermalization.

The natural quantity to characterize the equilibrium regime is the (asymptotic) time average that for site probabilities is defined as

$$\overline{\rho(j_0 | j, t)} := \lim_{T \to \infty} \frac{1}{T} \int_0^T dt \rho(j_0 | j, t) \tag{6}$$

Such a distribution, which is not influenced by the initial transient whose contribution vanishes by performing the asymptotic time average, is shown as a black trace in the last two panels of Fig. 1 and, with a better resolution in Fig. 3a. It represents the equilibrium average of the site populations. Notice that the computation of the time average is not explicitly required for the calculation of $\overline{\rho(j_0 | j, t)}$. Because of the lack of degeneration in the energy spectrum of the Hamiltonian Eq. (1), one can employ the relation $\overline{\exp\{i(E_k - E_{k'})t/\hbar\}} = \delta_{k,k'}$ in the time averaging, so deriving on the ground of Eqs. (5) and (6) the relation

$$\overline{\rho(j_0 \mid j,t)} = \sum_{k=1}^{N} |\langle j_0 | E_k \rangle|^2 |\langle j | E_k \rangle|^2 \tag{7}$$

requiring only the Hamiltonian eigenstates.

Despite the time averaging, the distribution $\overline{\rho(j_0 \mid j,t)}$ displays a kind of roughness in its site dependence (see the histogram in Fig. 3a), which is due to the local effects of the randomly chosen site energies. In a realization of the site energies, each site is differentiated at random from the neighbors, and this prevents the derivation of a smooth profile for the site dependence. On the other hand, each realization of the site energy is statistically equivalent. Therefore, in order to deal with the general properties of the model Hamiltonian Eq. (1), one should consider the average over the disorder realizations, as it is typically done in random matrix theory, or in the studies on Anderson localization where particular attention is paid to the self-averaging properties of observables.[25] In our treatment, the average over the disorder realizations can be conveniently done by averaging on the initial localization $j_0$ of the excitation. Because of the periodicity of the tight-binding chain, this is equivalent to rotate the given sequence of site energies for a given initial localization of the excitation, so averaging the local effects of specific values of site energies. In this way, only the diagonalization of a single system Hamiltonian is required while by using a direct sampling of the site energies, each realization of the Hamiltonian would require a separate diagonalization. Given the eigenvectors of the Hamiltonian, Eq. (7) is used to evaluate the time averaged probability for all the possible initial localizations $j_0$ in order to calculate the corresponding average for a given site displacement $\Delta j := j - j_0$

$$\left\{\overline{\rho(j_0 \mid j,t)}\right\} := \frac{1}{N}\sum_{j_0=1}^{N} \overline{\rho(j_0 \mid j_0 + \Delta j, t)} = \frac{1}{N}\sum_{j_0=1}^{N}\sum_{k=1}^{N} |\langle j_0 | E_k \rangle|^2 |\langle j_0 + \Delta j | E_k \rangle|^2 \tag{8}$$

In this way we recover a simple profile for the site probabilities as shown by the red line in Fig. 3a, with a smooth decay by moving away from the initial localization site. It represents the appropriate distribution to describe the localization attained in the equilibrium regime within the chain. It should be mentioned, however, that the average of Eq. (8) is strictly equivalent to the full average on the realizations of site energies only in the limit of an infinite $N \to \infty$ chain. For finite but large chain lengths, as in the cases reported here, a weak dependence on the site energy realizations can still be detected.

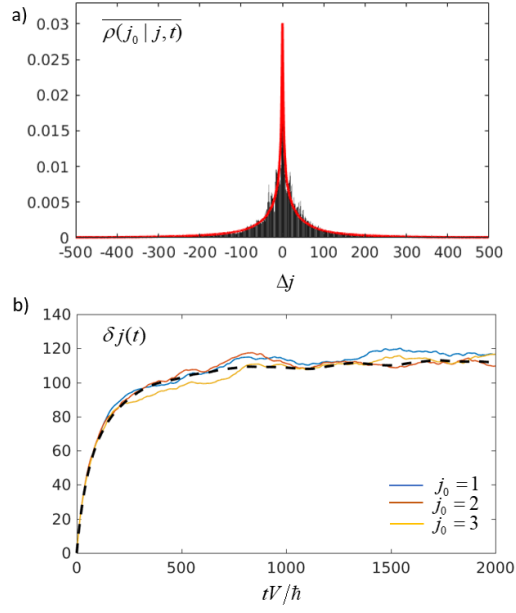

**Figure 3**. a) time averaged probability $\overline{\rho(j_0 \mid j,t)}$ as a function of the site displacement $\Delta j = j - j_0$ in the same conditions of Figures 1-2. The red line denotes the distribution $\left\{\overline{\rho(j_0 \mid j,t)}\right\}$, obtained by further averaging over N site $j_0$ of possible initial localization. b) time dependence of the square root of the Mean Squared Site Displacement, $\delta j(t)$ in Eq. (9), for three initial conditions $j_0 = 1, 2, 3$. The result of its average according to Eq. (10) with respect to the initial localization site is reported with a dashed line.

The increase with the time of the width of site distribution $\rho(j_0 \mid j,t)$ is the most evident feature emerging from the data of Fig. 1. It can be quantified by examining the Mean Squared Site Displacement (MSSD) defined as

$$\delta j(t)^2 := \sum_{\Delta j=-N/2}^{N/2-1} \Delta j^2 \rho(j_0 \mid j_0 + \Delta j, t) \qquad (9)$$

where $\Delta j = j - j_0$ is the site displacement constrained to the domain $-N/2 \le \Delta j < N/2$ taking advantage of the cyclic property of site indices. Its time dependence is displayed in Fig. 3b for three different initial sites $j_0$ of localization. Their different time dependence is evident, even if they share a similar trend. Because of chain periodicity, these differences can be interpreted again as an effect of realization of site energies. In order to remove it, we introduce like in Eq. (8) the average with respect the initial site localization $j_0$

$$\left\{\delta j(t)^2\right\} = \frac{1}{N} \sum_{j_0=1}^{N} \sum_{\Delta j=-N/2}^{N/2-1} \Delta j^2 \rho(j_0 \mid j_0 + \Delta j, t) \qquad (10)$$

whose square root is represented in Fig. 3b by a dashed line. Even in this case, the average over the initial states allows the partial averaging on the site energy disorder, which becomes complete in the limit of an infinite chain. In conclusion, the MSSD defined in Eq. (10) supplies a direct quantification of localization by looking at the time evolution of site populations arising from an initial fully localized state. Of particular interest is the equilibrium localization length defined as the square root of $\left\{\overline{\delta j(t)^2}\right\}$ obtained from Eq. (10) by replacing the site distribution $\rho(j_0 | j_0 + \Delta j, t)$ with its time average $\overline{\rho(j_0 | j_0 + \Delta j, t)}$ (or equivalently by taking the average with respect to the initial localization $j_0$ of the time averaged $\overline{\delta j(t)^2}$ mean squared displacement). It describes the localization of the site probability density in the equilibrium regime independently of the realization of site energies and, therefore, it depends only on the intrinsic parameters of the Hamiltonian model: the disorder strength $\sigma_\varepsilon / V$ and the length $N$ of the tight-binding chain.

## 3. Quantifying localization

The mean equilibrium squared site displacement (MSSD) introduced in the previous section quantifies the localization on the basis of the equilibrium average distribution on the sites of the chain. The extent of the localization is controlled by the chain disorder: let us first consider the upper limit of disorder strength, $\sigma_\varepsilon / V \to \infty$, when the hopping process is ineffective because of the off-diagonal elements in the Hamiltonian Eq. (1) can be neglected. In this condition the site probabilities do not change in time and the site displacement vanishes, $\forall t: \rho(j_0 | j, t) = \delta_{j,j_0}$, $\delta j(t) = 0$. The system is fully localized at the initial state and we do not observe any delocalization dynamics. In the other limit of vanishing disorder strength, $\sigma_\varepsilon / V \to 0$, all the sites become equivalent and the equilibrium regime corresponds to a uniform probability density, $\overline{\rho(j_0 | j, t)} = 1/N$. In this limit, the equilibrium MSSD attains its maximum value, $\overline{\delta j(t)^2} = (N^2 + 2)/12$ denoting the complete delocalization of the excitation within the periodic chain. Intermediate values of $\left\{\overline{\delta j(t)^2}\right\}$ are obtained from the numerical calculations with finite values of the disorder strength, and Fig. 4a shows the dependence of the square root of the MSSD (dots) on the disorder strength $\sigma_\varepsilon / V$. The increasing of the disorder leads to more localized equilibrium density profiles. Moreover, in Fig. 4 we report the MSSD for chains of two different lengths, $N = 10^3$ (blue dots) and $N = 10^4$ (red dots). As long as the MSSD is much shorter than the chain length $N$, further increases of the chain length do not modify the localization profile, meaning that the same localization length also characterizes the infinite chain. Conversely, finite size effects due to the chain length are clearly visible in the weak disorder limit.

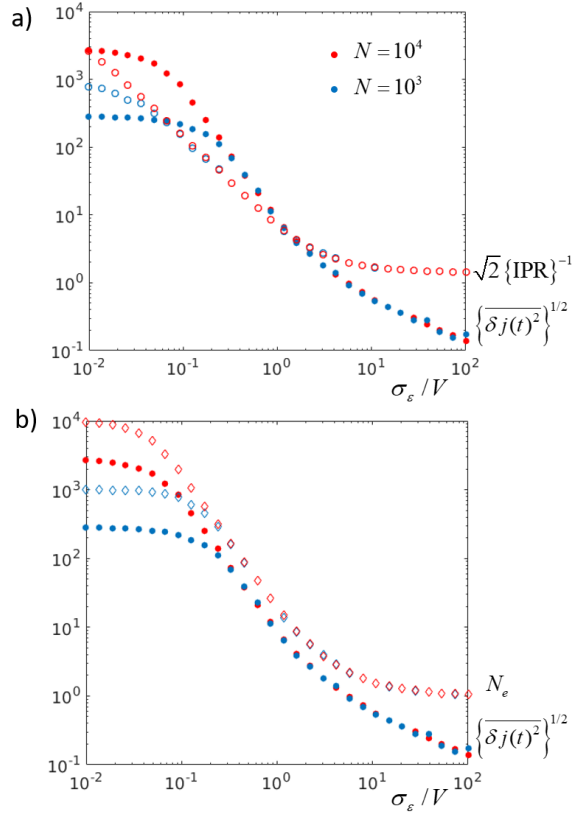

*Figure 4. Time average mean squared site displacement (MSSD) averaged over the initial sites (filled circles) as a function of the disorder strength $\sigma_\varepsilon/V$ for two different chain lengths as reported in the figure. a) the MSSD is compared with the inverse of the average IPR (empty circles), Eq. (11), b) the MSSD is compared with the effective number of states $N_e$ (diamonds) as determined by the entropic measure, eq. (14).*

It is well known that the phenomenon of localization is closely related to the statistical properties of the energy spectrum[34] and of the eigenstates[35, 36] of the disordered system. Intuitively, eigenfunctions delocalized over the sites are associated to non-vanishing electronic transport and a localization length is typically defined as the inverse of the exponent characterizing the fall-off of the eigenstate components in the site basis.[25] A general formula for the localization length in the 1D Anderson model is still not known, but expressions for the limit cases of strong and weak disorder have been derived (see e.g. refs [37, 38] and reference therein). A closely related quantity, the Inverse Participation Ratio (IPR) was often employed to quantify the localization of an eigenstate in the analysis of Anderson localization,[25, 36, 39] $\mathrm{IPR}(k) := \sum_{j=1}^{N} |\langle j|E_k\rangle|^4$. The inverse of the IPR is a measure of how many localized basis $|j\rangle$ "participate" to the $k$-th eigenstate $|E_k\rangle$. In this form, the IPR is a quantity that refers to each single eigenstate and it is not trivially related to the equilibrium average profile attained by the site population density which is determined by many energy components as specified by eq. (7). It is therefore of interest to analyze the connection between the IPR and the MSSD as different quantifiers of localization. To this aim, let us introduce the mean Inverse Participation Ratio $\{\mathrm{IPR}\}$ by averaging $\mathrm{IPR}(k)$ over all the eigenstates:

$$\{\text{IPR}\} := \frac{1}{N}\sum_{k=1}^{N}\text{IPR}(k) = \frac{1}{N}\sum_{k=1}^{N}\sum_{j=1}^{N}\left|\langle j|E_k\rangle\right|^4 \tag{11}$$

Its value quantifies the localization of the eigenvectors on average. The maximum value, $\{\text{IPR}\}=1$, corresponds to complete localization which is attained at $\sigma_\varepsilon/V \to \infty$ when each eigenstate corresponds to a chain site, $|E_k\rangle = |j\rangle$. In the opposite limit of complete delocalization, when $\sigma_\varepsilon/V \to 0$, each eigenvector is equally distributed on all the sites, $|\langle j|E_k\rangle| = 1/\sqrt{N}$ and the average IPR attains its smallest value $\{\text{IPR}\}=1/N$. Therefore, $\{\text{IPR}\}$ can be taken as the parameter quantifying the average localization from the point of view of the site distribution of the eigenstates.

The inverse of eq. (11), the participation ratio $\{\text{IPR}\}^{-1}$, was also interpreted as an average effective number of quantum states that significantly contribute to the quantum dynamics.[21, 22] Indeed, by comparing eq. (11) with the site and time average of the conditional probability Eq. (8) for $\Delta j = 0$ we can understand how the statistical properties of the eigenvalues are connected to the long time average of the system dynamics

$$\left\{\overline{\rho(j_0|j_0,t)}\right\} = \frac{1}{N}\sum_{j_0=1}^{N}\sum_{k=1}^{N}\left|\langle j_0|E_k\rangle\right|^4 = \{\text{IPR}\} \tag{12}$$

The above equality says that the mean IPR is equivalent to the mean time-averaged survival probability of the initial state. Moreover, the conditional probability is normalized as $\sum_j \rho(j_0|j,t) = 1$, and by approximating this normalization as the product of its maximum at $j = j_0$, eq.(12), and its width proportional to the square root of $\left\{\overline{\delta j(t)^2}\right\}$, one obtains the following approximate relation between the MSSD and the average participation ratio

$$\left\{\overline{\delta j(t)^2}\right\}^{1/2} \approx \sqrt{2}\{\text{IPR}\}^{-1} \tag{13}$$

Fig. 4a shows both the r.h.s. and the l.h.s. of eq. (13) in the entire range of disorder strength $\sigma_\varepsilon/V$ for two different chain lengths $N$. It should be mention that deviations between the localization amplitude and the participation ratio are expected both in the low disorder limit, because of finite size effects, and in the strong disorder limit, because eq. (13) is based on a continuous approximation. However, Fig. 4a also points out that both quantifiers in eq.(13) display a linear regime in the log-log representation for intermediate range of disorder strengths but with different slopes. Such discrepancy is not easily explained and points to a non-trivial dependence of the equilibrium average density profile, eq. (8), on the disorder strength,[40, 41] an issue that we will not study further here. Nonetheless, the different behavior of the mean square displacement and the participation ratio can be understood on an intuitive ground by considering that the former is sensitive to the whole distribution profile while the latter effectively measures only its maximum value.

Another parameter that has been extensively used to measure the portion of phase space accessed by a molecular system during its quantum dynamics is an entropy-like measure.[23, 42] The effective number $N_e$ of localized quantum states $|j\rangle$ that are dynamically connected to the initial state $|j_0\rangle$ can be defined as

$$\ln N_e = -\left\{\sum_j \overline{\rho(j_0|j,t)} \ln \overline{\rho(j_0|j,t)}\right\} \tag{14}$$

where we have also introduced the average over the initial state $j_0$ on the r.h.s. of Eq.(14). In Fig. 4b, the effective number of states derived from the entropy measure is compared with the average MSSD of the equilibrium density profile and they show the same dependence on the disorder strength in the intermediate regime. In conclusion, we showed that the characterization of localization based on the equilibrium average density profiles is related to standard statistical properties of the eigenstates as described by the IPR parameter. On the other hand, the equilibrium average distribution on the chain sites reflects the measure of the phase space accessed by the system during its quantum dynamics, and such a measure is well described by the entropy function of Eq.(14).

### 4. Fluctuations from delocalized random pure states

In the previous sections, the confinement due to the disorder of site energies was directly observed by looking at the time evolution of an initially localized state. On the experimental side, this requires the complete control on the quantum pure state at a given time. Depending on the physical system that is described by the model Hamiltonian of eq. (1), the control on the initial state may be achieved by site selective excitation or application of a strong site confining potential that can be then removed to initiate the particle equilibration dynamics. In this section, we analyze the Anderson localization from the opposite perspective of the lack of control or knowledge of the initial conditions. More specifically, we suppose that the initial pure state is chosen at random within the Hilbert space. Then, the initial state $|\psi(0)\rangle$ is already delocalized over all the chain and the site probabilities $|\langle j|\psi(t)\rangle|^2$, whose time evolution was examined in the section 2, do not display any form of confinement. The following issue then arises: is it possible to recognize the effects of Anderson localization also from the evolution of randomly chosen pure states and which observable can be used to detect them? The aim of the following analysis is that of providing a precise answer to this question.

Let us first introduce the random sampling of the initial pure state $|\psi(0)\rangle$. This is equivalent, in the framework of pure-state quantum statistical mechanics,[9, 43] to define the statistical ensemble of pure states of an isolated quantum system at equilibrium. Such an issue has been examined in the past from different points of view with the purpose of providing well defined rules for the random sampling of pure states within the Bloch hypersphere.[44-47] In the following, we shall employ the Random Pure State Ensemble which has been proposed and analyzed in detail elsewhere.[12, 31] The initial state $|\psi(0)\rangle$ is parameterized according to the set of phases $\alpha := (\alpha_1, \alpha_2, \cdots, \alpha_N)$ and the set of populations $P := (P_1, P_2, \cdots, P_N)$ with respect to eigenstates $|E_k\rangle$ of Eq.(3)

$$|\psi(0)\rangle = \sum_{k=1}^{N} \sqrt{P_k} e^{i\alpha_k} |E_k\rangle \qquad (15)$$

The uniform distribution on the Bloch hypersphere of the pure states is recovered from the uniform distribution on $N$-dimensional torus for the phases and the uniform distribution on $(N-1)$-dimensional simplex for the eigenstate populations taking into account the constraints of normalization, $\sum_{k=1}^{N} P_k = 1$, and positivity, $P_k \geq 0, \forall k$. At the computational level, this corresponds to an independent sampling of each phase according to the homogeneous distribution in the domain $[0, 2\pi)$, while for the population an efficient algorithm based on the independent sampling of $(N-1)$ variables in the domain $[0,1)$ is available as reported in ref[26]. It should be mentioned that RPS Ensemble can be used to describe different thermal conditions by restricting the pure state to subspaces of the full Hilbert space of the system, and this allows the evaluation of the thermodynamic properties of the quantum system.[31] Here we consider the simplest application of RPSE statistics without any restriction on the Hilbert space, which corresponds to the infinite temperature limit.

Once the initial conditions are specified according to the previous parameterization of $|\psi(0)\rangle$, the time dependence of the pure state $|\psi(t)\rangle$ is easily determined by replacing in Eq. (15) the initial values of the phases $\alpha_k(0) \equiv \alpha_k$ with their time dependent counterparts $\alpha_k(t) := \alpha_k(0) - E_k t/\hbar$, the populations $P$ being constants of motion. The site probabilities are given by

$$\rho_j(t) := |\langle j|\psi(t)\rangle|^2 = \langle \psi(t)|(|j\rangle\langle j|)|\psi(t)\rangle \qquad (16)$$

That is the expectation value of the site projection operator $|j\rangle\langle j|$. Unlike the site probability Eq. (5) introduced in Section 2, these site populations do not depend anymore on the site $j_0$ of initial localization because the initial random pure state is in general a superposition of the site basis. Nonetheless, the site probabilities at any time still depend on the choice of the initial wave-function $|\psi(0)\rangle$, that is on the sets of eigenstate populations $P$ and of initial phases $\alpha(0)$. In general, the site probabilities $\rho_j(t)$ display a fluctuating time dependence very much like that of Fig. 2 for the time evolution of an initially localized state. However, their distribution on the sites of the chain is completely different. This emerges quite clearly by considering the asymptotic time average as in Eq. (6) to obtain

$$\overline{\rho_j(t)} := \lim_{T \to \infty} \frac{1}{T} \int_0^T dt \rho_j(t) = \sum_{k=1}^{N} P_k |\langle j|E_k\rangle|^2 \qquad (17)$$

with an explicit dependence on the eigenstate populations $P$ only, because the time average eliminates the phases. In the upper panel of Fig. 5 we have represented these average site populations for a realization of the initial pure state $|\psi(0)\rangle$ according to RPSE statistics, with the same Hamiltonian employed for the calculations of Figs. 1-3 of Section 2. By comparison with the upper panel of Fig. 3, the difference with respect to the localization process becomes evident. While the time average $\overline{\rho(j_0|j,t)}$ of site probabilities considered in Section 2 leads to a picture of localization according to the confinement in a finite domain centered on the initial site $j_0$, the time evolution of an initial random pure state $|\psi(0)\rangle$ does not produce any evident form of localization. On the contrary, the time averaged site probabilities of Eq. (17), which are shown in Fig. 5a, display a random site dependence along the chain, which is controlled by the realization of the population set $P$. With another set of populations, different values of time averaged probabilities would be obtained, but with an analogous random dependence on the site.

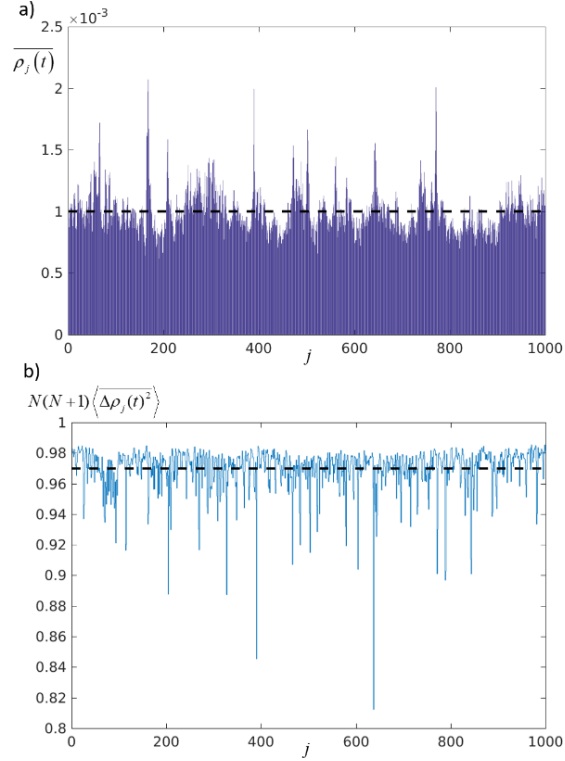

**Figure 5**. a) Site $j$ dependence of the time averaged populations $\overline{\rho_j(t)}$ for a random realization of the initial pure state. Their ensemble average $\langle \overline{\rho_j(t)} \rangle = 1/N$ is represented with a dashed line. b) Fluctuation amplitudes $\langle \overline{\Delta\rho_j(t)^2} \rangle$ of site probability Eq.(20) scaled by the dimensional factor $N(N+1)$. The dashed line denotes the average Eq. (21) over the sites of the chain. The results are for a chain of dimension $N = 10^3$ and disorder strength $\sigma_\varepsilon / V = 0.24$.

To estimate general properties of random pure states that are functions of the population set, $f(P)$, the average with respect to the distribution on the eigenstate populations can be introduced:

$$\langle f(P) \rangle := \int dP \, p_{RPSE}(P) f(P) \tag{18}$$

where $p_{RPSE}(P)$ is the RPSE probability density on the set $P$ of populations.[12] The corresponding average of Eq. (17) is given as

$$\left\langle \overline{\rho_j(t)} \right\rangle = \frac{1}{N} \sum_{k=1}^{N} \langle j | E_k \rangle \langle E_k | j \rangle = \frac{1}{N} \qquad (19)$$

because $\sum_k |E_k\rangle\langle E_k|$ is a decomposition of unity and the average populations are $\langle P_k \rangle = 1/N$ because of the statistical equivalence of the different populations in the RPSE. In conclusion, constant site probabilities are recovered when the average of Eq. (19) is employed to analyze the results independently of the eigenstate population realization. This clearly points out that the site populations deriving from a random initial pure state do not display evident forms of localization, in opposition to the confinement found from the evolution of the initially localized state as shown in the first panel of Fig. 3.

In order to find whether the site energy disorder has any influence on the evolution of a random initial pure state, one has to examine the statistical moments of order higher than the simple time average. Their formal treatment in the framework of RPSE statistics has been presented elsewhere[32] by considering the expectation values of an operator, and it can be applied to the present case by taking into account that site population $\rho_j(t)$ Eq. (16) is the expectation value of the site projection operator $|j\rangle\langle j|$. The amplitude of its fluctuations around the equilibrium average $\Delta\rho_j(t) := \rho_j(t) - \overline{\rho_j(t)}$ in the time domain, is described by the time average of its square $\overline{\Delta\rho_j(t)^2}$ which in general depends on the set $P$ of eigenstate populations. A population independent estimate of the fluctuation amplitude is given by its average $\left\langle \overline{\Delta\rho_j(t)^2} \right\rangle$ on the ensemble of random pure states. According to the RPSE statistics, the following simple relation is recovered

$$\left\langle \overline{\Delta\rho_j(t)^2} \right\rangle = \frac{1}{N(N+1)} \left( 1 - \sum_{k=1}^{N} |\langle j | E_k \rangle|^4 \right) \qquad (20)$$

which is a special case of the general result reported in Eq. (22) of ref[32]. It allows a direct connection between the fluctuation amplitude and the spreading of the site basis $|j\rangle$ on the eigenstate $|E_k\rangle$ basis. Indeed, one can easily recognize two limits: a vanishing fluctuation amplitude if $|j\rangle$ is exactly one eigenstate, while the amplitude of fluctuations $\left\langle \overline{\Delta\rho_j(t)^2} \right\rangle$ attains its maximum value of about $1/N^2$ if the components of $|j\rangle$ are nearly homogeneously distributed amongst the eigenvectors, that is if $|\langle E_k | j \rangle| = 1/\sqrt{N}$. The fluctuation amplitude as defined in Eq. (20) is independent of the realization of initial pure state, but it still depends on the specific site location $j$. In the bottom panel of Fig. 5 we have represented the amplitude of site probability fluctuations as a function of the site location. Such an irregular dependence is due to the random character of the tight-binding Hamiltonian that determines fluctuations of different amplitude according to the local realizations of site energies. In order to deal with a general result for the model system independently of its particular realization, the average on the possible site energies would be required. However, like for Eq.(8) in the previous analysis of the evolution of an initially localized state, we employ the computationally more convenient short cut of the average on the sites of the chain, that for the fluctuation amplitude of site probabilities is defined as

$$\left\{\left\langle\overline{\Delta\rho_j(t)^2}\right\rangle\right\} := \frac{1}{N}\sum_{j=1}^{N}\left\langle\overline{\Delta\rho_j(t)^2}\right\rangle = \frac{1}{N(N+1)}\left(1 - \frac{1}{N}\sum_{j=1}^{N}\sum_{k=1}^{N}\left|\langle j|E_k\rangle\right|^4\right) \qquad (21)$$

With reference to fluctuation amplitudes of all the chain sites which are reported in Fig. 5b, it represents simply their mean and it is drawn as a black dashed line. Furthermore, by recalling Eq. (11), one can specify the fluctuation amplitude $\left\{\left\langle\overline{\Delta\rho_j(t)^2}\right\rangle\right\}$ according to the mean Inverse Participation Ratio

$$\left\{\left\langle\overline{\Delta\rho_j(t)^2}\right\rangle\right\} = \frac{1-\{\text{IPR}\}}{N(N+1)} \qquad (22)$$

This is an important result because it leads to a straightforward and somehow unexpected relation between fluctuation amplitude of random pure states and the localization quantified by the parameter $\{\text{IPR}\}$. It implies that physical conditions leading to an increase of the localization, that is a stronger disorder of site energies, produces a decrease of fluctuation amplitude for a given chain length $N$. Our analysis clearly demonstrates that the localization of eigenstates, as described by parameter $\{\text{IPR}\}$, controls the fluctuations of site probabilities during the evolution of a random pure state. On the other hand, in the previous section we have shown that $\{\text{IPR}\}$ and, therefore, fluctuation amplitude of random pure state, is related to the localization parameters deriving from the evolution of an initially localized state, like the localization length attained at equilibrium or the effective number of states involved in the quantum dynamics as given by the entropic measure of Eq. (14). Completely different dynamics for the random pure states or for initially localized states supply complementary information about Anderson localization. Furthermore, we mention that the analysis of fluctuations does not have only a conceptual value. As a matter of fact, the fluctuations of expectation values or, equivalently, the fluctuations of elements of reduced density matrix, have been characterized in recent experiments with confined ultra-cold ions.[6] This opens the possibility of observing localization effects by experimental measures of fluctuations.

In general, each site population $\rho_j(t)$ has a twofold variability due to its time dependence, in the one hand, and the dependence on the randomly chosen initial pure state $|\psi(0)\rangle$, on the other hand. The previous analysis quantifies the time variability according to the fluctuation amplitude. In order to characterize the other kind of variability, let us consider the site population equilibrium average $\overline{\rho_j} \equiv \overline{\rho_j(t)}$. Eq. (17) specifies its dependence on the eigenstate populations $P$ and if a diagram like the top panel of Fig. (5) is examined for a different realization of these populations, then a different profile for the site dependence of $\overline{\rho_j}$ would be observed but with a similar random behavior. We can therefore evaluate the variance of the equilibrium site distribution within the ensemble of random pure state. By recalling Eq. (18) for the average with respect to the eigenstate population distribution, such a variance is defined as

$$\sigma_j^2 := \left\langle \left( \overline{\rho_j} - \left\langle \overline{\rho_j} \right\rangle \right)^2 \right\rangle \tag{23}$$

It quantifies the variability of the time average $\overline{\rho_j}$ that derives from its eigenstate population dependence. By specializing the RPSE general result Eq. (21) in ref[32] to the projection operator $|j\rangle\langle j|$, one derives the explicit relation

$$\sigma_j^2 = \frac{1}{N(N+1)} \left( \sum_{k=1}^{N} |\langle j|E_k\rangle|^4 - \frac{1}{N} \right) \tag{24}$$

The comparison of this result for the ensemble variance with Eq. (20) for the fluctuation amplitude leads to the conclusion that they are not independent. The two kinds of variance are complementary, in the meaning that they satisfy the following relation

$$\left\langle \overline{\Delta \rho_j(t)^2} \right\rangle + \sigma_j^2 = \frac{N-1}{N^2(N+1)} \tag{25}$$

that is, their sum depends only on the Hilbert space dimension $N$. This implies that if we change the system Hamiltonian, for instance by modifying the energy disorder in the tight-binding Hamiltonian Eq. (1), each of the two variances would change in general, but with the constraint Eq. (25) of a constant sum. Such a "circle law" holds for each site $j$ for a given realization of the energy disorder. By performing like in Eq. (21) the average on the chain sites, one obtains a form the circle law which is independent of the realization of site energy disorder in the limit of an infinite chain

$$\left\{ \left\langle \overline{\Delta \rho_j(t)^2} \right\rangle \right\} + \left\{ \sigma_j^2 \right\} = \frac{N-1}{N^2(N+1)} \tag{26}$$

where the first term at the l.h.s. is given by eq. (22) while the ensemble variance $\{\sigma_j^2\}$ averaged over the chain sites takes the explicit form

$$\{\sigma_j^2\} = \frac{1}{N(N+1)} \left( \{\text{IPR}\} - \frac{1}{N} \right) \tag{27}$$

Fig. 6 shows the circle law in the form given by Eq. (26). The two axes correspond to the two contributions to the total variance of the site populations, namely the ensemble variance as the abscissa and the fluctuation amplitude as the ordinate. The different points in the circle are the results obtained for different values of the disorder strength $\sigma_\varepsilon/V$: in the limit of strong disorder $\sigma_\varepsilon/V \to \infty$, when the system remains strongly localized in a relaxation experiment, time fluctuations are suppressed while the equilibrium site populations show the maximum of the variance for different realization of random pure states. In the opposite limit of vanishing disorder $\sigma_\varepsilon/V \to 0$, when the equilibrium site distribution is delocalized over the finite chain, one can observe the maximum amplitude of site population fluctuations due to the dynamics of the wave-function while time averaged site populations show a vanishing variance for different realization of the random pure state. A similar interplay between fluctuation amplitude and variances of quantum observable in the microcanonical ensemble was analyzed in ref. [29] where the vanishing of the ensemble average eq. (24) in the thermodynamic limit was associated to the integrability to chaos transition and the emergence of ergodicity. Here we have shown that in finite systems the relative weight of fluctuations and ensemble variance can be associated to the extent of localization.

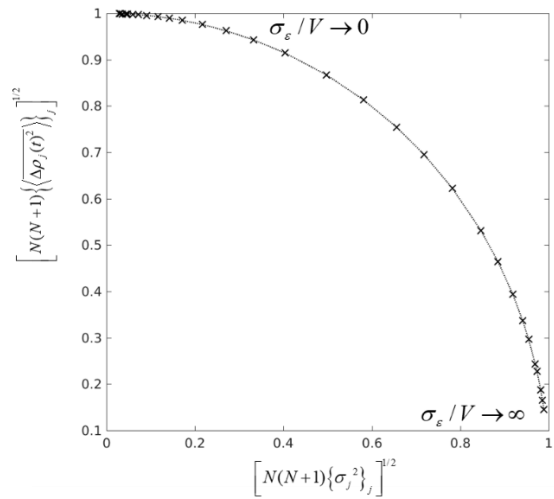

*Figure 6. The circle law Eq.(26) which relates the fluctuation amplitude and the ensemble variance of site population averaged over the disorder. The crosses represent different value of the disorder strength $\sigma_\varepsilon/V$ and the two limits discussed in the text are also pointed out in the figure.*

5. Conclusion

The dynamics of the simplest Anderson tight-binding Hamiltonian for different disorder strengths was investigated to capture different manifestations of quantum localization. We have analysed both the confinement of the quantum state resulting from the time evolution of an initially localized state and the dynamics of random pure states which are intrinsically delocalized. In the former case, the concept of dynamical equilibration plays a central role, by allowing a straightforward analysis of the disorder induced localization according to the time averaged site occupation probabilities. Furthermore, as shown in section 3, such an approach allows to establish the connection with other quantifiers of Anderson localization, like the inverse participation ratio and the effective number of states involved in the quantum dynamics as given by the entropic measure. On the other hand, the use of quantum statistical ensembles leads to a direct characterization of random pure states in terms of fluctuation amplitude and of the variability with respect to the initial conditions. Both the amplitude of the equilibrium site distribution resulting from the evolution of an initially localized state (MSSD) and the amplitude of the fluctuations of random pure states are determined by the statistical properties of the transformation matrix between the site representation and the eigenstates. Therefore, the central result that emerges from our analysis is a clear connection between the process of equilibration observed when the system is initialized in a non-typical state (the fully localized initial state) and the amplitude of dynamical fluctuations in typical random pure state. It is rather intriguing that such a connection, having a thermodynamic flavour, emerges in a system that does not thermalize. Moreover, we showed that the fluctuation amplitude of the site population due to the underlying unitary dynamics and their variance associated to the uncertainty of the quantum state are related by the "circle law" that is shown in Fig. 6 and both reflect the extent of the localization.

The one-dimensional tight-binding model with diagonal disorder is the simplest model of Anderson localization. This simplicity allowed us to carry out a methodologically clear analysis that only relies on the microscopic picture provided by quantum mechanics with no additional postulates. Nonetheless, we believe that the resulting picture is rather general and the extension to other systems should be possible. For example, the problem of Anderson localization serves as a simple model for investigating the dynamics of vibrational or electronic excitons in molecular and inorganic crystals, and for understanding the quantum mechanical energy flow in complex molecules.[48, 49] Important generalizations of this work include the study of many-body interacting system for which thermalization would not occur[50-52] and multidimensional system.

**Acknowledgements**

BF acknowledges the support of the Italian Ministero dell'Istruzione, Università e Ricerca through the grant Rita Levi Montalcini (2013). All authors acknowledge support by the DOR funding scheme of the Dipartimento di Scienze Chimiche, Universitá degli Studi di Padova.